\documentclass[11pt,a4paper]{article}
\usepackage[margin=25mm]{geometry}

\usepackage{amsmath,amssymb,graphicx,booktabs,array,natbib}
\usepackage{algorithm,algpseudocode}
\usepackage[hidelinks]{hyperref}
\hypersetup{pdftitle={Representation-split Carleman lattice Boltzmann with near-unit step success probability on quantum computers},pdfauthor={Hua-Dong Yao, Sauro Succi}}
\graphicspath{{figures/}}
\newcommand{\Ma}{\mathrm{Ma}}
\newcommand{\Lc}{\mathcal{L}}
\newcommand{\Qc}{\mathcal{Q}}
\newcommand{\Kc}{\mathcal{K}}
\newcommand{\Cc}{\mathcal{C}}
\title{Representation-split Carleman lattice Boltzmann with near-unit step success probability on quantum computers}
\author{Hua-Dong Yao$^{1}$\thanks{huadong.yao@chalmers.se}\hspace{0.35em} and Sauro Succi$^{2}$\\[4pt] \small $^1$Chalmers University of Technology, Gothenburg, Sweden\\ \small $^2$Fondazione Istituto Italiano di Tecnologia,\\ \small Center for Life Nano-Neuroscience at La Sapienza, Rome, Italy}
\date{September 2026}
\begin{document}
\maketitle

\begin{abstract}
The Carleman lattice Boltzmann method replaces the quadratic collision of the lattice Boltzmann equation by a linear map on a lifted state. Existing block encodings of this map succeed with probability $10^{-5}$ to $10^{-2}$ per step, so that the probability of a multi-step evolution decays exponentially with the number of steps. This study develops multi-relaxation-time Carleman lattice Boltzmann, a representation split in which the Carleman-lifted collision is performed in a moment basis on the amplitude-encoded state, while streaming remains a permutation in population space. The level-2 hierarchy closes under collision but not under streaming, which generates the work of pressure, advection and stress on the momentum flux. The state therefore remains the full pair array, and the moment representation is an orthogonal change of basis of the population one, with identical dynamics and truncation error. The operator to be block encoded changes with the basis, and with it the success probability of a quantum step. In the weighted Hermite basis with the amplitude encoding $f_i/\sqrt{w_i}$, the relaxation is a diagonal operator of norm one, and the level-2 collision stage becomes this diagonal plus a sparse coupling from products of conserved moments. A structured block encoding of this form raises the single-step success probability on the D2Q9 lattice from $10^{-2}$ for the best population encoding to 0.98. With the pair sector in relative coordinates and scaled, one step requires $O(\log^2 N)$ Toffoli gates and is verified by statevector simulation. The encoding and readout overhead of a $T$-step run is then polynomial in $T$ and $N$ rather than accumulating a factor $p^T$. For resolved flows, the remaining exponential loss in coherent multi-step Carleman lattice Boltzmann evolution is set by the non-equilibrium norm relaxed by dissipation rather than by the block encoding.
\end{abstract}

\section{Introduction}
Quantum computers hold the prospect of information processing capabilities with no known efficient classical counterpart \citep{NielsenChuang2010,Preskill2018}. A pure state of $n$ qubits is described by $2^n$ complex amplitudes, so that a quantum circuit acts on a state space of exponential dimension. The practical realization of this capability struggles against classicalization, the tendency of quantum systems to behave classically at sufficiently large scales and temperatures \citep{SucciSanavioLove2026}. Coherence and entanglement, the resources on which quantum computation rests, are fragile against environmental perturbations, so that present devices run circuits of limited depth before errors accumulate \citep{Preskill2018}. Fault-tolerant architectures can suppress logical errors at a substantial overhead in physical qubits and in the cost of non-Clifford gates \citep{Fowler2012}. These constraints determine the quantities by which a quantum algorithm is assessed, namely the number of logical qubits, the number of non-Clifford gates per time step and the depth of the circuit that has to be run coherently. Moreover, an amplitude-encoded state holds the field in the amplitudes of the state vector. A non-unitary linear map of such a state is implemented as a block of a larger unitary and recovered by postselection, which succeeds with a probability below one \citep{GSLW2019}. A failed postselection discards the run, so that a low success probability multiplies the number of repetitions and exposes the full depth of the circuit to noise in each of them.

The difficulty is sharper for classical fluid dynamics. The coherent evolution of the quantum circuit model is linear and unitary, whereas the equations of fluid dynamics are nonlinear and dissipative. \citet{SucciSanavioLove2026} identify nonlinearity and dissipation as the two obstacles that a quantum algorithm for classical fluids has to address at the level of the mathematical formulation, before any circuit is designed. One route to representing the nonlinearity is to embed the system into a larger linear one. Carleman linearization maps a polynomial system onto an infinite linear hierarchy of tensor powers and truncates, and \citet{Liu2021} showed that the truncation converges when dissipation dominates nonlinearity, which for a viscous flow amounts to a bound on the Reynolds number. \citet{Gaitan2020} applied a quantum algorithm for ordinary differential equations to the flow through a convergent-divergent nozzle. \citet{Bharadwaj2023} developed a hybrid quantum linear-systems workflow at gate level for Poiseuille and Couette flows, with quantum postprocessing of nonlinear functions of the velocity field, as a step toward the quantum simulation of nonlinear flows. Dissipation enters as a non-unitary operator and is subject to the block-encoding cost described above. \citet{SucciEPL2023} review the state of the field and its prospects for fluid dynamics.

The lattice Boltzmann (LB) equation is a natural starting point for the quantum route, because its two substeps separate the difficulties. Streaming is a permutation of the populations and hence unitary, while the nonlinearity and the dissipation are both confined to the local collision, the former through the equilibrium and the latter through the relaxation. \citet{Mezzacapo2015} proposed a quantum simulator of the LB equation on pseudospin-boson platforms, in which the non-unitary collision is realized by a heralded protocol. \citet{Todorova2020} constructed quantum circuits for the collisionless streaming step, and \citet{Budinski2021} built a full circuit for the advection-diffusion equation, whose collision is linear. Nonlinear flows require a linear representation of the quadratic collision, which \citet{Wang2025} obtain from a node-level ensemble description of a lattice gas with the linear collision rule of lattice gas cellular automata. Within a Carleman lift the lifted collision remains a non-unitary operator, so that a central quantum cost of an LB step is the block encoding of the lifted collision and its success probability.

Applied to the LB equation, the Carleman lift replaces the populations $f$ by the tensor powers $(f, f\otimes f, \dots)$ and truncates, so that one time step becomes a linear map on the lifted state \citep{ItaniSucci2022}. The resulting Carleman lattice Boltzmann (CLB) method, truncated at the second tensor power, reproduces the LB dynamics of a two-dimensional Kolmogorov flow at Reynolds numbers between 10 and 100, with a root-mean-square error below $10^{-3}$ over a hundred steps at the lower Reynolds numbers \citep{SanavioSucci2024}. The obstacle to a quantum implementation is the block encoding of the lifted collision. With sparse-access oracles the D2Q9 relaxation matrix has sparsity 81 and succeeds with probability $6\times10^{-5}$ per step \citep{SanavioSimonSucci2025}, so that direct multi-step evolution is impractical without amplification. \citet{BastidaZamora2026} keep the Carleman variables in population space, $f$ and $f\otimes f$ with Bhatnagar-Gross-Krook (BGK) collision, and store the pair sector in coordinates relative to the first site, which makes the collision the same operator at every site and brings the per-step cost to $O(Q^3 + \log^2N)$, with $Q$ the number of velocities. Their circuit runs several steps at a per-step success probability of order $10^{-2}$, which depends on the state and on the relaxation time. The total success probability still decays as $p^T$, which they attribute to the non-unitarity of the collision rather than to the encoding. \citet{Jennings2026} identify the convergence of the Carleman linearization, the time stepping, the scaling of the condition number and the data extraction as the end-to-end bottlenecks, and find that a modest quantum advantage may be preserved for selected observables in a high-error-tolerance regime. The three Carleman routes of \citet{SanavioThreeRoutes2024} and the perspective of \citet{SucciEPL2023} place these constructions in context.

Two complementary approaches proposed in 2026 address the same object, the diagonal multi-relaxation-time (MRT) relaxation of the moments, $\delta m_k' = (1-\omega_k)\,\delta m_k$. \citet{KhanSucciYao2026} realize it deterministically as a completely positive trace-preserving (CPTP) map on a two-rail expectation-level encoding, with success probability one for the dissipative substage. Streaming remains classical, and the nonlinear equilibrium is left to a Carleman lift in a hybrid pipeline (their Sec.~5.4). The second approach, multi-relaxation-time Carleman lattice Boltzmann (MRT-CLB), was first outlined in an internal note \citep{SucciNote2026}. Classical MRT schemes already collide in moment space and stream in population space \citep{LallemandLuo2000}. Applied to the Carleman-lifted operator on the coherent, amplitude-encoded state, this factorization makes the lifted relaxation diagonal in both the level-1 and the pair sector, the quadratic coupling sparse, and streaming a permutation. The relative-coordinate encoding of \citet{BastidaZamora2026} makes the collision local in the site registers but leaves it a dense operator in population space.

Despite the body of work reviewed above, the structure of the Carleman-lifted MRT collision in moment space, and the success probability of its block encoding on an amplitude-encoded state, have not been examined. The present study develops MRT-CLB along the lines of the note. The arrangement, collision in moment space and streaming in population space applied to the lifted state, is referred to below as the representation split. The representation split makes the collision diagonal and sparse, whereas the relative-coordinate encoding makes it local, and the present study combines the two devices. The diagonal structure is expected to raise the success probability. The quadratic sector of the moment hierarchy might also close at the second level with twelve components per site, the nine moments and three products of the conserved momentum. Three questions are examined, namely whether the moment hierarchy remains closed at the second level once streaming is included, how the diagonal moment-space collision affects the single-step success probability, and what circuit resources result. The resulting algorithm is also compared with the CPTP approach.

The contribution of this paper is threefold. First, the level-2 hierarchy in the moment representation closes under collision but not under streaming, which generates cross-site pressure, advection and stress-work terms outside the twelve-component local state. The state remains the full pair array, and the moment lift is an orthogonal change of basis of level-2 population Carleman with the same dynamics and truncation error (Section~\ref{sec:closure}). Second, in the $w$-weighted Hermite basis with the weighted amplitude encoding the collision stage is a diagonal plus a sparsity-three coupling. Its single-step success probability rises from the $10^{-5}$ class of the sparse-access population construction to $0.11$ with the same construction. With a structured encoding it reaches $0.98$, compared with $10^{-2}$ for the optimal population-space encoding. The dependence on the rates, the viscosity and the Mach number is a few percent for resolved flows. The analysis attributes the low success probability of the existing constructions to the Euclidean metric of the encoding and to the density of the population basis. The dissipation itself removes only the non-equilibrium norm relaxed in the step (Section~\ref{sec:structure}). Third, with the pair sector stored in the relative coordinates of \citet{BastidaZamora2026}, one step is an explicit circuit on $4\log_2 L+14$ qubits with $O(\log^2 N)$ Toffoli gates, counted and verified by simulation. With the pair sector scaled, the encoding and readout overhead of a $T$-step run is polynomial rather than exponential in the number of steps (Section~\ref{sec:circuit}). The level-2 lift and its truncation are taken from \citet{SanavioSucci2024}, and the relative-coordinate encoding from \citet{BastidaZamora2026}. The closure analysis, the moment-space Carleman formulation with its weighted encoding, the pair-sector scaling and the counted circuit are new. Section~\ref{sec:setting} defines the model, the bases and the lift, Section~\ref{sec:limits} states the range of validity of the results, and Section~\ref{sec:conclusion} concludes.

\section{Lattice Boltzmann, moment bases and the level-2 lift}\label{sec:setting}
The D2Q9 lattice is used in lattice units, with velocities $c_i$ ($i = 0,\dots,8$), weights $w_i$ and sound speed $c_s^2 = 1/3$, on a periodic box of $N = L^2$ sites. The populations $f_i(x,t)$ define the density $\rho = \sum_i f_i$ and the momentum $J = \sum_i c_i f_i$, and collision relaxes them towards the equilibrium
\begin{equation}
f_i^{\rm eq} = w_i\left[\rho + 3\,c_i\!\cdot\! J + \tfrac92 (c_i\!\cdot\! J)^2 - \tfrac32 J\!\cdot\! J\right],
\label{eq:feq}
\end{equation}
in which the density has been set to one inside the quadratic term, the weakly compressible form used by \citet{SanavioSucci2024} and \citet{BastidaZamora2026}. This form differs from the standard second-order equilibrium, which divides the quadratic term by $\rho$, by $O(\Ma^2(\rho-1)) = O(\Ma^4)$. It also makes the equilibrium exactly quadratic in the populations, $f^{\rm eq} = E_1 f + E_2 (f\otimes f)$, as the Carleman lift requires. The quadratic part contributes no mass and no momentum at its own site.

Collision is performed in moment space. With $M$ a $9\times9$ matrix whose rows define nine moments $m = Mf$, the first three being $\rho$, $J_x$, $J_y$ up to normalization, the MRT collision relaxes each moment at its own rate,
\begin{equation}
m_k' = (1-\omega_k)\,m_k + \omega_k\,m_k^{\rm eq}, \qquad \omega_k\in(0,2),
\label{eq:mrt}
\end{equation}
and leaves the conserved moments unchanged. The shear viscosity is $\nu = c_s^2(1/\omega_\nu - 1/2)$, with $\omega_\nu$ the rate of the two shear moments, $\omega_e$ that of the trace and $\omega_q$ and $\omega_\varepsilon$ those of the three ghost moments. Equal rates give the BGK model. Written back in population space, the per-site update and the lattice-wide step read
\begin{equation}
f' = Lf + Q(f\otimes f), \qquad F(t+1) = \Lc F + \Qc(F\otimes F),
\label{eq:step}
\end{equation}
with $L = M^{-1}(I-\Omega)M + M^{-1}\Omega M E_1$, $Q = M^{-1}\Omega M E_2$, $\Omega = \mathrm{diag}(\omega_k)$, $F\in\mathbb{R}^{9N}$ the population vector, $\Lc = S(I_N\otimes L)$, $S$ the streaming permutation $f_i(x + c_i, t+1) = f_i'(x,t)$, and $\Qc = S(I_N\otimes Q)$ acting on same-site pairs only.

Two moment bases are used. Gram-Schmidt orthonormalization of the monomials $1$, $c_x$, $c_y$, $c_x^2+c_y^2$, $c_x^2-c_y^2$, $c_xc_y$, $c_xc_y^2$, $c_x^2c_y$ and $c_x^2c_y^2$, in that order, in the Euclidean inner product on $\mathbb{R}^9$ gives the basis of \citet{LallemandLuo2000} with normalized rows. For this basis $M = T$ is orthogonal, so that the population-to-moment map is a unitary on an amplitude-encoded state. The same process in the weighted inner product $\langle a, b\rangle_w = \sum_i w_i a_i b_i$ gives the tensor Hermite basis of D2Q9 up to a rotation of the trace and deviator, $M = H$ with $HWH^{\rm T} = I$, $W = \mathrm{diag}(w)$. In this basis the linear part of the equilibrium map is $HE_1H^{-1} = \mathrm{diag}(1,1,1,0,\dots,0)$, so that the level-1 collision block is the diagonal
\begin{equation}
C_1 = \mathrm{diag}(1,1,1,1-\omega_3,\dots,1-\omega_8),
\label{eq:c1}
\end{equation}
and the quadratic part of the equilibrium enters only the three second-order moments. $H$ itself is not orthogonal, but the map from the weighted populations $g_i = f_i/\sqrt{w_i}$ to the moments, $U = HW^{1/2}$, is orthogonal, and the rest state $g_i = \sqrt{w_i}$ has unit norm. The pair $(H, g)$ is referred to as the Hermite representation with weighted encoding, and $(T, f)$ as the Euclidean representation with plain encoding. The two bases define the same MRT scheme when all rates are equal and differ in the ghost sector otherwise, because the fourth-order Euclidean moment mixes the fourth-order Hermite moment with the trace. The Hermite MRT scheme is used throughout, and the Euclidean basis enters only as another representation of the same operator.

Carleman linearization lifts $F$ to $(F, F\otimes F, \dots, F^{\otimes K})$ and truncates at $K$. At level two the state $(F, F\otimes F)$ has dimension $9N + 81N^2$, the full tensor product rather than the symmetric pairs, so that operator sparsity multiplies. One step is then the block upper triangular matrix
\begin{equation}
\Cc = \begin{pmatrix}\Lc & \Qc\\ 0 & \Lc\otimes\Lc\end{pmatrix}
= \mathcal{S}\,\mathcal{T}^{\rm T}\Kc\,\mathcal{T}, \qquad \mathcal{S} = S\oplus(S\otimes S), \qquad \mathcal{T} = (I_N\otimes U)\oplus(I_N\otimes U)^{\otimes 2}.
\label{eq:lift}
\end{equation}
The population sector is exact after one step, because the update is quadratic. The truncation error enters at the second step, at $O(\Ma^3)$ in the populations, because the dropped term is cubic in the fluctuations. In the factorized form $\mathcal{S}$ is a permutation, $\mathcal{T}$ is orthogonal, and $\Kc$ is the lifted collision in moment space, block upper triangular with $I_N\otimes C_1$ and $(I_N\otimes C_1)^{\otimes2}$ on the diagonal and the quadratic coupling above it. This factorization, the representation split, is the MRT-CLB algorithm. Two properties of the lift are used below. Because $(\Lc\otimes\Lc)(G\otimes G) = (\Lc G)\otimes(\Lc G)$, the pair sector of a product initial state remains a product, and the level-2 dynamics reduces to
\begin{equation}
F(t+1) = \Lc F(t) + \Qc\big(G(t)\otimes G(t)\big), \qquad G(t) = \Lc^t F_0,
\label{eq:product}
\end{equation}
with $G(t)$ the linear LB evolution of the initial state. This makes level-2 trajectories on lattices of any size computationally inexpensive. The pair sector may also be scaled, $\psi_\lambda = (F, \lambda F\otimes F)$, which divides $\Qc$ by $\lambda$ in Eq.~(\ref{eq:lift}) and leaves the dynamics unchanged.

For a matrix $A$ with row and column sparsities $s_r$, $s_c$ (largest numbers of nonzeros), the sparse-access block encoding of \citet{GSLW2019} has the subnormalization $\alpha_s$, the optimal block encoding $\alpha_2$, and on a state $\psi$ the single-step success probability is
\begin{equation}
\alpha_s = \sqrt{s_rs_c}\,\max_{ij}|a_{ij}|, \qquad \alpha_2 = \|A\|_2, \qquad p = \frac{\|A\psi\|^2}{\alpha^2\|\psi\|^2} = \frac{r}{\alpha^2},
\label{eq:p}
\end{equation}
with $r = \|A\psi\|^2/\|\psi\|^2$ the norm ratio and $r_\lambda$ its value on the scaled state $\psi_\lambda$. Errors of a truncated evolution against the nonlinear reference are measured on the velocity and on the populations relative to their flow part,
\begin{equation}
e_u = \frac{\|u - u_{\rm ref}\|}{\|u_{\rm ref}\|}, \qquad e_f = \frac{\|F - F_{\rm ref}\|}{\|F_{\rm ref} - F_{\rm rest}\|}, \qquad e_q = \frac{\|q - JJ^{\rm T}\|}{\|JJ^{\rm T}\|},
\label{eq:errors}
\end{equation}
where $e_q$ is the error of a local quadratic sector $q$ against the reference. Two periodic test flows are used. The Taylor-Green vortex on $L\times L$ has
\begin{equation}
u_x = U_0\sin\kappa x\cos\kappa y, \qquad u_y = -U_0\cos\kappa x\sin\kappa y, \qquad \rho = 1 + \frac{U_0^2}{4c_s^2}(\cos2\kappa x + \cos2\kappa y),
\label{eq:tg}
\end{equation}
with $\kappa = 2\pi/L$ and equilibrium populations. Its kinetic energy decays as $e^{-4\nu\kappa^2t}$, and $t_{\rm visc} = 1/(4\nu\kappa^2)$ denotes this decay time. The Kolmogorov-like flow of \citet{SanavioSucci2024} has
\begin{equation}
f_i = w_i\left[1 + A_x\cos(2\pi k_xy/L)\,c_{i,x} + A_y\cos(2\pi k_yx/L)\,c_{i,y}\right], \qquad k_x = 1, \quad k_y = 4,
\label{eq:kolm}
\end{equation}
on the $32\times32$ lattice, with $A_y = 2A_x/3$ and $\kappa = 2\pi/L$. The Mach number is $\Ma = U_0/c_s$, with $U_0 = A_x/3$ for the Kolmogorov flow, so that the amplitudes $A_x = 0.3$, $A_y = 0.2$ of that paper correspond to $\Ma = 0.17$.

\section{Closure properties of the level-2 moment representation}\label{sec:closure}
This section examines whether the quadratic sector of the moment hierarchy can be closed locally or requires the pair array. A local formulation keeps, per site, the nine moments and the three products $q = JJ^{\rm T} = (J_x^2, J_xJ_y, J_y^2)$ of the conserved momentum, twelve components, because the quadratic part of the equilibrium depends on $q$ alone and $q$ is collision invariant. This construction closes exactly for the collision step, in line with the exactness of the level-2 lift for collision-only updates found by \citet{ItaniSucci2022} on the D1Q3 lattice. With $\rho = 1$ inside the quadratic term the per-site update of $(m, q)$ is the linear map with blocks $C_1$, $\Omega B$ and $I_3$, where $Bq$ is the quadratic part of the equilibrium moments, and this was verified to machine precision. To determine whether the closure extends to the streaming step, the quadratic sector after one full step is examined next.

After streaming, the momentum at $x$ and the quadratic sector are
\begin{equation}
J(x,t+1) = \sum_i c_i\, f_i'(x - c_i), \qquad q(x,t+1) = J(x,t+1)\,J(x,t+1)^{\rm T},
\label{eq:qstream}
\end{equation}
a linear combination of the post-collision moments of the nine neighbors and a bilinear form in them. To identify the terms generated by streaming, $q(x,t+1)$ was expanded exactly with bookkeeping parameters for the orders in the Mach number, in the non-equilibrium moments and in the gradient. The parameters are $J = \Ma\,u$, $m_s = (Bq)_s + \Ma\,\delta\,n_s$ for the second-order moments and $m_h = \Ma\,\delta\,h$ for the ghosts, and a factor $\kappa$ for every difference between a neighbor value and the value at $x$. The result has 12314 monomials in fifteen classes according to the kinds of factors they contain. Only the class of same-site products $u_au_b$, forty monomials, lies in the span of the local state $\{m(y), q(y)\}$. Table~\ref{tab:classes} lists the classes that enter at first order in the gradient with the continuum form of their leading terms. The remaining ten classes are products of two transport terms and enter at second order.

\begin{table}[t]\centering\small
\caption{Classes of monomials contributing to $q(x,t+1)$ through first order in the gradient, with their orders in $(\Ma, \delta)$, same-site and cross-site counts, and leading continuum forms. $\sigma'$ is the post-collision non-equilibrium stress.}\label{tab:classes}
\begin{tabular}{llccl}\toprule
class & order & same site & cross site & continuum form\\\midrule
$u\cdot u$ & $(2,0)$ & 40 & 126 & none at first order: linear transport gives $J + O(\kappa^2)$\\
$r\cdot u$ & $(3,0)$ & 28 & 164 & $2\,\mathrm{Sym}\,J\otimes(-c_s^2\nabla\rho)$, pressure work\\
$u\cdot u\cdot u$ & $(3,0)$ & 56 & 416 & $2\,\mathrm{Sym}\,J\otimes(-\nabla\!\cdot\!(JJ))$, advective flux\\
$n\cdot u$ & $(2,1)$ & 112 & 656 & $2\,\mathrm{Sym}\,J\otimes(\nabla\!\cdot\!\sigma')$, stress work\\
$h\cdot u$ & $(2,1)$ & 160 & 832 & flux of the ghost moments into $J$\\\bottomrule
\end{tabular}\end{table}

The continuum forms are the terms of
\begin{equation}
\frac{\mathrm{d}(JJ^{\rm T})}{\mathrm{d}t} = 2\,\mathrm{Sym}\big(J\,\partial_tJ^{\rm T}\big), \qquad \partial_tJ = -c_s^2\nabla\rho - \nabla\!\cdot\!(JJ) + \nabla\!\cdot\!\sigma .
\label{eq:qdot}
\end{equation}
At first order in the gradient the quadratic sector evolves by the work of the pressure gradient, of the advective flux and of the viscous stress on the momentum flux. Each of these terms is a product of $J$ at one site with a difference of a conserved or a non-conserved moment across sites, and none has an image in the local span. The transport of $q$ itself is the exception. The momentum flux of the linear equilibrium is pure pressure, so the linear transport of $J$ is $J + O(\kappa^2)$ and the $u\cdot u$ class has no first-order term. Accordingly, any local rule that reduces to the identity on uniform fields transports $q$ to second order. Two such rules were tested, the identity rule $q(x,t+1) = q(x)$ and the stencil
\begin{equation}
q(x,t+1) = \sum_i 3w_i\,\mathrm{Sym}\big(c_ic_i^{\rm T}\,q(x - c_i)\big),
\label{eq:stencil}
\end{equation}
the analogue of the momentum transport. On a smooth field their residuals decrease as $\kappa^{1.98}$ under lattice refinement. For lattice-scale roughness they are first order, with the leading first-order discrepancy given by the antisymmetric part of $J\otimes\nabla J$.

\begin{figure}[t]\centering\includegraphics[width=\textwidth]{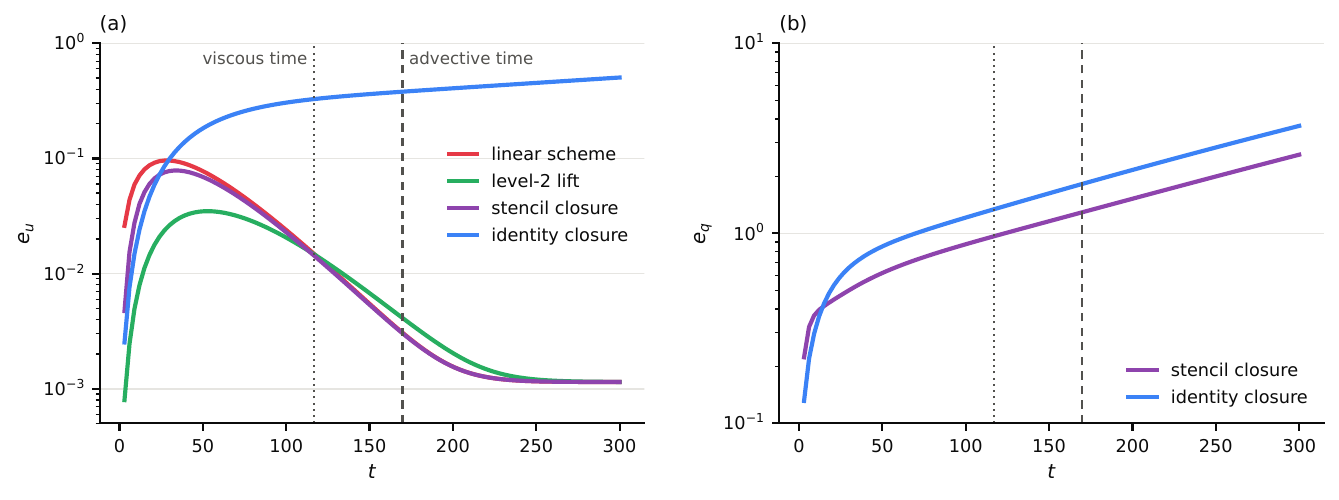}
\caption{The local closures depart from the pair-space reference on the viscous time scale. Kolmogorov flow on the $32\times32$ lattice with the amplitude reduced to $\Ma = 0.052$, $\omega = 1.5$. (a) Velocity error $e_u$ against the nonlinear reference, as a function of the time step $t$, for the linear scheme, the level-2 population lift and the two local closures (identity and stencil); the viscous (dotted) and advective (dashed) times are marked in both panels. (b) Error $e_q$ of the local quadratic sector against $JJ^{\rm T}$ of the reference.}\label{fig:closure}\end{figure}

Figure~\ref{fig:closure} shows the behavior of local schemes built on these rules. On the Kolmogorov flow of Eq.~(\ref{eq:kolm}) on the $32\times32$ lattice, with the amplitude reduced to $\Ma = 0.052$ and $\omega = 1.5$, the population level-2 lift has $e_u = 8\times10^{-4}$ at $t = 3$ compared with $2.6\times10^{-2}$ for the linear scheme, and is five to thirty times more accurate over the first twenty steps. The two errors meet near the viscous time and decay together to $1.1\times10^{-3}$ after the advective time $1/(\kappa U_0) = 170$. The two local schemes have the same moment-sector update, exact for one step, and differ only in $q$. The identity closure retains the initial quadratic-stress contribution after the flow has decayed, and its velocity error reaches 0.5 by $t = 300$. The stencil rule diffuses the spatial variation of $q$ at the lattice rate $1/6$ rather than the viscous rate, so that within tens of steps only the lattice mean of $q$ remains, which the stencil conserves and which exerts no force. The scheme then reduces to the linear one, whose velocity error it reproduces to three digits at all late times. In both schemes the quadratic sector departs from $JJ^{\rm T}$ on the viscous time scale (Fig.~\ref{fig:closure}b), and the populations acquire a pressure contribution that departs from the reference, $e_f = 0.07$ to $0.5$ compared with $10^{-3}$ for the level-2 lift. On the Taylor-Green vortex the pressure work and the advective flux cancel, and only the viscous decay of $q$ remains. The velocity is then unaffected, because the quadratic stress is a gradient, but the identity closure reaches $e_q = e^2 - 1 = 6.4$ at $t = 2t_{\rm visc}$, its undecayed value, and $e_f = 0.2$.

The cross-site pairs contain the work terms of the momentum flux, and the pair array stays. With the full pair array retained, the moment-space lift is the population lift in another orthogonal basis, $\mathcal{T}\Cc\mathcal{T}^{\rm T}$. It has identical dynamics, identical truncation error and the same validity window of level-2 truncation, the advective time or the viscous decay before it. For the $\omega = 1.9$, $\Ma = 0.17$ case considered here, the level-2 lift does not remain stable over 300 steps on the $32\times32$ lattice, and the moment representation shows the same behavior, as expected from the basis equivalence. The twelve local components remain an exact collision block within the full pair representation, and this block is where the moment formulation provides its computational advantage.

\section{The collision stage in the moment representation}\label{sec:structure}
\begin{figure}[t]\centering\includegraphics[width=\textwidth]{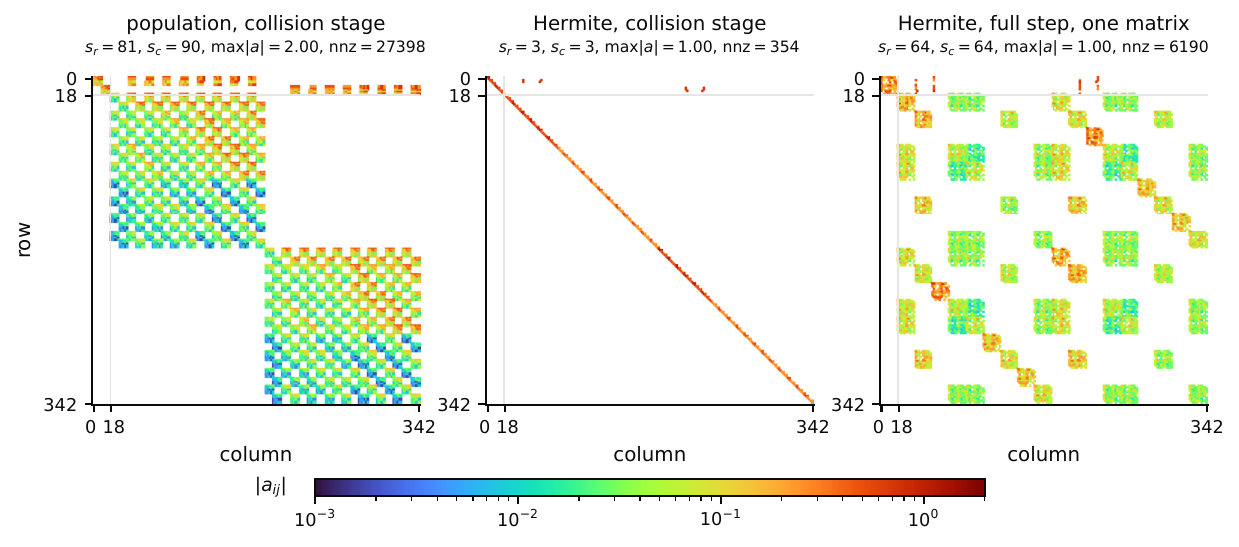}
\caption{Sparsity and magnitude of the level-2 collision stage on a two-site lattice (dimension 342, BGK $\omega=1.5$).}\label{fig:sparsity}\end{figure}

This section examines the effect of the change of basis on the collision operator. Table~\ref{tab:structure} gives the level-2 collision stage $\Kc$ of the Hermite MRT scheme in four representations on a two-site lattice. Every entry is independent of the lattice size, because the stage is block diagonal over site pairs. In the population representation with plain encoding, the construction of \citet{SanavioSimonSucci2025}, the level-1 block $L$ is dense, with nine nonzeros per row. The stage then has sparsities 81 and 90, and the quadratic coupling $Q$ has entries up to 2.0 at $\omega = 1.5$, so that $\alpha_s = 171$. Its singular values span 0.05 to 9.4, because the Euclidean metric on populations is not the one in which relaxation is a contraction. The weighted encoding $g = f/\sqrt w$ alone moves the singular values into $[0.25, 1.5]$ and reduces $\alpha_s$ to 28.5, a factor of six that is due to the metric. The Hermite basis reduces the sparsity to three in rows and columns and the largest entry to one, $\alpha_s = 3$ (Fig.~\ref{fig:sparsity}). Its Pauli decomposition, the sum of tensor products of Pauli matrices that a linear combination of unitaries (LCU) would use, fills 2.5 percent of all strings on two sites compared with 89 percent in the population representation, with a one-norm of the coefficients of 16 compared with 401 (Table~\ref{tab:structure}). The metric and the basis thus contribute separately to the reduction, the metric through the singular values and the basis through the sparsity.

\begin{table}[t]\centering\small
\caption{Level-2 collision stage of the Hermite MRT scheme for BGK relaxation at $\omega = 1.5$, compared in four representations. The Pauli filling and the one-norm $\alpha_P$ of the Pauli coefficients are evaluated on the two-site lattice. The remaining quantities are independent of the lattice size.}\label{tab:structure}
\begin{tabular}{lcccccccc}\toprule
representation & $s_r$ & $s_c$ & $\max|a|$ & $\alpha_s$ & $\sigma_{\max}$ & $\sigma_{\min}$ & Pauli filling & $\alpha_P$\\\midrule
population, plain encoding & 81 & 90 & 2.00 & 171 & 9.39 & 0.052 & 0.89 & 401\\
population, weighted encoding & 81 & 90 & 0.33 & 28.5 & 1.50 & 0.25 & 0.68 & 112\\
moment, Euclidean basis, plain & 4 & 9 & 4.50 & 27 & 9.39 & 0.052 & 0.18 & 190\\
moment, Hermite, weighted & 3 & 3 & 1.00 & 3 & 1.50 & 0.25 & 0.025 & 15.8\\\bottomrule
\end{tabular}\end{table}

These numbers follow from the structure of the stage. The level-1 block $C_1$ of Eq.~(\ref{eq:c1}) is diagonal with entries of modulus at most one, and the pair block $C_1\otimes C_1$ is diagonal with products of these entries. The only off-diagonal contribution is the quadratic coupling from the pair sector into the level-1 sector, per site the $3\times4$ block
\begin{equation}
C_{\rm site} = \tfrac12\,\mathrm{diag}(\omega_e, \omega_\nu, \omega_\nu)\begin{pmatrix}1 & 1 & 0 & 0\\ 1 & -1 & 0 & 0\\ 0 & 0 & 1 & 1\end{pmatrix}
\label{eq:csite}
\end{equation}
from the pair components $(m_1m_1, m_2m_2, m_1m_2, m_2m_1)$ to the trace, the deviator and the shear moment, with singular values $\omega_e/\sqrt2$ and $\omega_\nu/\sqrt2$ (twice). The norm 1.5 of the stage in Table~\ref{tab:structure} is due to this coupling alone. With the pair sector scaled by $\lambda$ the coupling is divided by $\lambda$, and $\|\Kc_\lambda\|_2$ falls to 1.077, 1.007 and 1.001 at $\lambda = 3$, 10 and 30. The full step written as one matrix, on the other hand, is denser in moment space than in population space, because streaming couples each moment to the nine moments of nine neighbors, with $s = 59^2$ on $3\times3$ and $4\times4$ compared with 81 and 90 in population space. The gain is confined to the factorized step of Eq.~(\ref{eq:lift}), permutation streaming, orthogonal transform and sparse collision, which is the MRT-CLB prescription in quantitative form.

\begin{figure}[t]\centering\includegraphics[width=\textwidth]{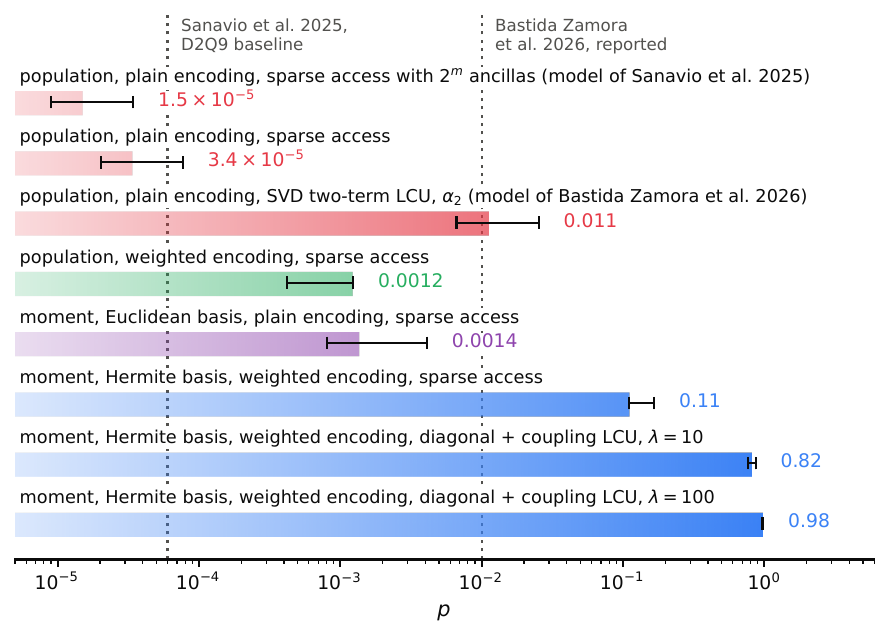}
\caption{Single-step success probability $p$ of the collision stage by representation and block-encoding construction, on the $32\times32$ Taylor-Green state at $\Ma = 0.087$ and $\omega = 1.5$ (bars, whose values are printed at the right); the horizontal line through each bar spans the range of $p$ over $\omega = 1.0$ to $1.95$, with viscosity $\nu = (1/\omega - 1/2)/3$. The fill of each bar darkens with $p$ on the same scale in every row, reaching full opacity at $p = 1$. The dotted vertical lines, annotated in the panel, are the D2Q9 baseline $6\times10^{-5}$ of \citet{SanavioSimonSucci2025} and the value of order $10^{-2}$ reported by \citet{BastidaZamora2026} at $N = 16$, whose construction is modeled as the optimal subnormalization of the population-space stage.}\label{fig:success}\end{figure}

The norm ratio $r$ of the collision stage was computed along level-2 trajectories of the Taylor-Green vortex on $16\times16$ and $32\times32$ at $\Ma = 0.02$ to $0.2$, with BGK rates from 1.0 to 1.95 and three MRT sets with ghost rates from 1.05 to 1.95. The Kolmogorov flow at $\Ma = 0.17$ was included as well. In the weighted representation $r$ has its minimum, 0.9992, at the second step of the Taylor-Green runs. The minimum is 0.9989 when the initial state carries a 30 percent non-equilibrium perturbation (a pseudo-random component with zero mass and momentum at every site, of norm 0.3 times that of the flow part), and 0.9972 through the transient of the Kolmogorov flow. The success probability is set by the subnormalization, and Fig.~\ref{fig:success} compares the constructions. With the sparse-access encoding it is 0.111 for the Hermite stage at every rate but $\omega = 1$, where one nonzero per row vanishes and it is 0.167. In population space it is $2.0\times10^{-5}$ to $7.7\times10^{-5}$, and $p$ follows $1/\max|Q|^2$ with $\max|Q| = 4\omega/3$ and rises as $\omega$ falls, a dependence introduced by the encoding rather than by the relaxation. The per-step probability of \citet{BastidaZamora2026} depends on the relaxation time in the same way. Their two-term linear combination of the singular value decomposition is the block encoding with $\alpha_2 = \sigma_{\max}$ of the population-space stage. It gives $r/\sigma_{\max}^2 = 1.1\times10^{-2}$ at $\omega = 1.5$ (Table~\ref{tab:structure}) and $0.34$ at their $\omega = 0.2$ for a near-equilibrium state, a range that contains their reported value of order $10^{-2}$. The Hermite weighted stage has $\sigma_{\max} = 1.5$ at $\lambda = 1$ and approaches one as $\lambda$ grows. The present reconstruction of the sparse-access construction gives $3.4\times10^{-5}$ at $\omega = 1.5$, and $1.5\times10^{-5}$ with the subnormalization 256 of the seven-ancilla construction (Table~\ref{tab:resources}), compared with the $6\times10^{-5}$ reported by \citet{SanavioSimonSucci2025}.

Exploiting this structure further reduces the subnormalization relative to the generic sparse-access construction. The diagonal part has entries of modulus at most one and is block encoded with subnormalization one by a controlled rotation of one ancilla per moment register. The coupling is block encoded through its singular value decomposition, with subnormalization $\omega_{\max}/\sqrt2$, $\omega_{\max} = \max(\omega_e, \omega_\nu)$. Their linear combination has
\begin{equation}
\alpha = 1 + \frac{a}{\lambda}, \qquad a = \frac{\omega_{\max}}{\sqrt2}, \qquad p = \frac{r_\lambda}{(1 + a/\lambda)^2}.
\label{eq:alcu}
\end{equation}
At $\omega = 1.5$ this gives $p = 0.82$ at $\lambda = 10$ and $0.98$ at $\lambda = 100$, and across the rates 0.77 to 0.87 and 0.97 to 0.99. The optimal encoding, $\alpha_2 = \|\Kc_\lambda\|_2$, would give 0.44 at $\lambda = 1$ and approach $r$ at large $\lambda$.

To separate the physical contribution to $p = r/\alpha^2$ from that of the encoding, the norm lost in a collision is examined first. In the Hermite representation the diagonal part of the level-1 stage removes the norm fraction
\begin{equation}
1 - r^{(1)} = \sum_k \big[1 - (1-\omega_k)^2\big]\,\phi_k,
\label{eq:loss}
\end{equation}
with $\phi_k$ the norm fraction of moment $k$, and the pair sector about twice that. The coupling restores the equilibrium part of the second-order moments, so that the net loss is the non-equilibrium norm relaxed in the step, the discrete counterpart of the entropy production. On the trajectories above, the loss is concentrated in the shear sector. Its norm fraction is $2\times10^{-7}$ to $5\times10^{-4}$ on Taylor-Green, increasing with the Mach number, and $1.3\times10^{-3}$ at the peak of the Kolmogorov transient, while the bulk and ghost sectors are two to four orders of magnitude below. Changing the ghost rates from 1.05 to 1.95 changes $r$ by less than $3\times10^{-6}$ at any time on these trajectories, and changing the shear rate from 1.0 to 1.95, a factor of forty in the viscosity, changes it by less than $2\times10^{-4}$. Dissipation thus reduces the success probability only in proportion to the relaxed non-equilibrium norm, which is of order $(\mathrm{Kn}\,\Ma)^2$ for a resolved flow, with Kn the ratio of the relaxation length to the flow scale, and $r$ stays above 0.997 on every trajectory considered. The factor $1/\alpha^2$ carries the remainder, from $171^{-2}$ in the Euclidean population encoding to close to unity in the structured encoding of the weighted Hermite stage. Two causes of the low success probability of the existing constructions are thus identified, the Euclidean metric of the encoding, which is not the metric in which relaxation is a contraction, and the density of the population basis, whereas the dissipation itself is a minor third. The weighted encoding removes the first and the Hermite basis the second.

\section{Circuit and resources}\label{sec:circuit}
This section gives the resource count of one time step and of a run. One step of the factorized algorithm was built as a circuit and verified against the classical lift by statevector simulation. The encoding of the pair sector and its streaming follow \citet{BastidaZamora2026}. The collision stage, the transforms, the pair-sector scaling and the gate count are the part specific to the moment representation. The pair sector is stored in coordinates relative to the first site, $\psi[x, i, r, j] = \lambda X[(x,i),(x+r,j)]$. The same-site pairs on which the coupling acts then form the slice $r = 0$, so that the collision is the same operator at every site. The level-1 sector occupies the padding slot 15 of the second moment register, $\psi[x, i, 0, 15] = F[x,i]$. The padded transform, the streaming and the diagonal leave that slot invariant, so that no level qubit is needed (\citet{BastidaZamora2026} use an order register). Every stage is a tensor product over the two factors. The registers are the site $x$ and the relative position $r$, $2n$ qubits each for $L = 2^n$, two moment registers of four qubits, and six ancillas. The total is $4n + 14$ qubits, 22 on $4\times4$. One step consists of the stages listed in Algorithm~\ref{alg:step}. The collision stage, line 4, implements the linear combination of Eq.~(\ref{eq:alcu}). The diagonal block encoding consists of two uniformly controlled rotations on ancillas, and the block encoding of $C_{\rm site}$ on the slice $r = 0$ is one ancilla rotation between two small fixed rotations, with the flag $r = 0$ computed and uncomputed. The streaming, line 6, consists of controlled incrementers, the shift of \citet{BastidaZamora2026} written in relative coordinates. The state is prepared directly in the simulation, and its preparation is outside the count.

\begin{algorithm}[t]
\caption{One time step of the MRT-CLB circuit on the site register $x$, the relative-position register $r$, the moment registers $m_1$ and $m_2$, and six ancillas. Gate counts are Toffolis, from Eq.~(\ref{eq:gates}), unless stated otherwise.}\label{alg:step}
\begin{algorithmic}[1]
\State \textbf{Preparation} $\psi[x,i,r,j] \leftarrow \lambda X[(x,i),(x+r,j)]$, $\psi[x,i,0,15] \leftarrow F[x,i]$ \Comment{outside the count}
\For{$t = 1$ to $T$}
\State \textbf{Transform} $(U\oplus I_7)\otimes(U\oplus I_7)$ on $m_1$, $m_2$ \Comment{200 CNOT, generic bound}
\State \textbf{Collision} linear combination of Eq.~(\ref{eq:alcu}), $C_{\rm site}$ on the slice $r = 0$ \Comment{$4(2n-1) + 198$}
\State \textbf{Inverse transform} on $m_1$, $m_2$ \Comment{200 CNOT, generic bound}
\State \textbf{Streaming} $x \mathrel{+}= c_{i_1}$, $r \mathrel{+}= c_{i_2} - c_{i_1}$ \Comment{$36n^2 + 204n + 12$}
\State \textbf{Postselection} of the ancillas \Comment{$p = r_\lambda/\alpha^2$}
\EndFor
\State \textbf{Readout} of the level-1 sector with amplitude amplification \Comment{$O(\lambda\sqrt N)$ repetitions}
\end{algorithmic}
\end{algorithm}

Table~\ref{tab:verify} lists the verified steps on $4\times4$ (22 qubits, Taylor-Green at $\Ma = 0.17$). After $k$ steps the postselected state equals $\Cc_\lambda^k\psi/\alpha^k$ to $10^{-13}$ in both sectors. This was verified over three consecutive steps at $\lambda = 10$, for one step at $\lambda = 100$ and for one step with the MRT rates, the table listing the largest relative error over the two sectors. The success probability of each step, measured as the postselected norm fraction, equals the prediction $r_\lambda/\alpha^2$ from the classical state to four digits. The value 0.80 at the second step has a physical origin. On this coarse lattice, $\kappa = \pi/2$, the linear evolution generates 1.3 percent of non-equilibrium norm in one step, of which 1.0 percent is in the deviator. At $\omega = 1.5$ the level-1 sector relaxes three quarters of this norm and the pair sector twice that, 2.0 percent, which with $\alpha = 1.106$ gives the measured 0.80. On the $32\times32$ lattice at the same Mach number the non-equilibrium norm generated in one step is $4\times10^{-4}$.

\begin{table}[t]\centering\small
\caption{Statevector verification of the $4\times4$ circuit for the Taylor-Green flow at $\Ma = 0.17$. MRT rates are listed as $\omega_\nu/\omega_e/\omega_q/\omega_\varepsilon$.}\label{tab:verify}
\begin{tabular}{lccccc}\toprule
rates & $\lambda$ & step & rel.\ error & $p$ measured & $r_\lambda/\alpha^2$\\\midrule
BGK 1.5 & 10 & 1 & $3.2\times10^{-14}$ & 0.8171 & 0.8171\\
BGK 1.5 & 10 & 2 & $6.1\times10^{-14}$ & 0.8012 & 0.8012\\
BGK 1.5 & 10 & 3 & $9.1\times10^{-14}$ & 0.8133 & 0.8133\\
BGK 1.5 & 100 & 1 & $3.3\times10^{-14}$ & 0.9788 & 0.9788\\
MRT 1.3/1.6/1.1/1.8 & 10 & 1 & $3.2\times10^{-14}$ & 0.8068 & 0.8068\\\bottomrule
\end{tabular}\end{table}

\begin{figure}[t]\centering\includegraphics[width=\textwidth]{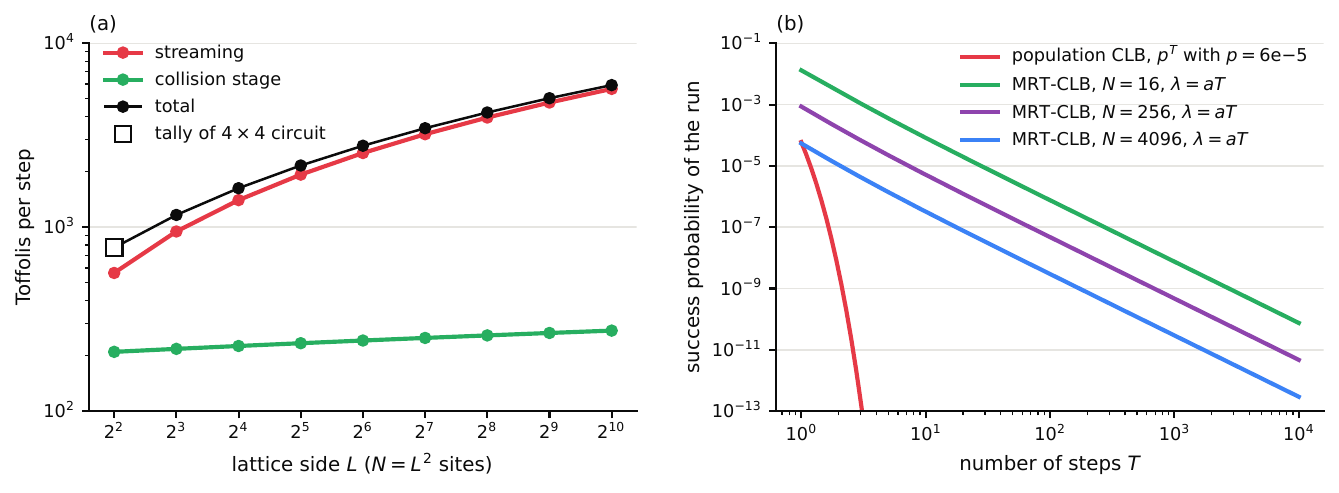}
\caption{(a) Toffoli gates per step against the lattice side for the streaming and collision stages, with the tally of the $4\times4$ circuit. (b) Success probability of a $T$-step run with readout of the nonlinear register at $\lambda = aT$, for three lattice sizes, against $p^T$ of the population construction.}\label{fig:resources}\end{figure}

The $4\times4$ circuit requires 774 Toffoli gates per step, 564 in the streaming and 210 in the collision stage. The four-qubit transforms and the diagonal rotations add 466 CNOT and 549 single-qubit gates. In this count a $k$-controlled X is $2(k-1)$ Toffolis, a uniformly controlled rotation on $m$ controls is $2^m$ CNOT, and each transform is taken at the generic bound of 100 CNOT. In general the Toffoli counts of the collision and streaming stages are
\begin{equation}
N_{\rm coll} = 4(2n-1) + 198, \qquad N_{\rm stream} = 36n^2 + 204n + 12,
\label{eq:gates}
\end{equation}
where the lattice enters the collision only through the flag $r = 0$ and the streaming consists of 36 controlled shifts of $n$ bits whose $j$-th bit carries four or five controls plus the $j$ lower bits. The transforms add 400 CNOT independent of $N$ (Fig.~\ref{fig:resources}a). Per step the count is $O(\log^2N)$, dominated by streaming. The same scaling holds for population Carleman, where streaming is also the quadratic part \citep{SanavioSimonSucci2025}, and for the local encoding of \citet{BastidaZamora2026}, where the collision is a dense population-space block decomposed generically at $O(Q^3)$ gates.

The cost of a run follows from the per-step probability of Eq.~(\ref{eq:alcu}) and from the readout. The level-1 register, which holds the nonlinear solution, has the norm fraction $\|F\|^2/(\|F\|^2 + \lambda^2\|G\|^4) = 1/(1 + \lambda^2N)$ in the weighted encoding, since $\|F_g\|^2 = N$ and $\|G_g\|^4 = N^2$ near rest. A $T$-step run followed by a measurement of that register succeeds, for $r = 1$, with
\begin{equation}
P = \frac{(1 + a/\lambda)^{-2T}}{1 + \lambda^2N}, \qquad \lambda^\ast = aT, \qquad P^\ast = \frac{e^{-2}}{a^2T^2N},
\label{eq:total}
\end{equation}
where $\lambda^\ast$ maximizes $P$ (Fig.~\ref{fig:resources}b). Amplitude amplification of the final readout replaces the factor $1/(\lambda^2N)$ by $O(\lambda\sqrt N)$ repetitions, so a $T$-step evolution requires $O(T^2\sqrt N\log^2N)$ gates. Table~\ref{tab:resources} compares the four constructions, with $\sigma_{\max}$ of the local population construction taken from Table~\ref{tab:structure} at $\omega = 1.5$ and from the same computation at $\omega = 0.2$, and the CPTP numbers from Sections 3 and 5.2 of \citet{KhanSucciYao2026}. Under the present scaling the encoding and readout overhead of a $T$-step run is polynomial in $T$ and $N$. The only remaining multiplicative factor is $\prod_t r_t$ of the relaxed non-equilibrium norm, with a loss of order $(\mathrm{Kn}\,\Ma)^2$ per step for a resolved flow. The population constructions accumulate the per-step factor as $p^T$, with $p = 6\times10^{-5}$ for the sparse-access encoding and of order $10^{-2}$ for the local encoding of \citet{BastidaZamora2026}. The CPTP approach is deterministic for the dissipative relaxation substage. This count does not establish an advantage over classical lattice Boltzmann, whose cost is $O(TN)$. It shows instead that the coherent Carleman route can be run over many steps.

\begin{table}[t]\centering\small
\caption{Four quantum constructions of the lattice Boltzmann collision stage and of a time step. Entries are those reported in the cited works or, where stated, our reconstruction or evaluation of their construction. The last row is evaluated at $r = 1$.}\label{tab:resources}
\begin{tabular}{>{\raggedright\arraybackslash}p{22mm}>{\raggedright\arraybackslash}p{28mm}>{\raggedright\arraybackslash}p{28mm}>{\raggedright\arraybackslash}p{30mm}>{\raggedright\arraybackslash}p{26mm}}\toprule
 & population CLB, sparse access \citep{SanavioSimonSucci2025} & population CLB, local \citep{BastidaZamora2026} & MRT-CLB (this work) & CPTP relaxation \citep{KhanSucciYao2026}\\\midrule
state & amplitude, $9N+81N^2$ & amplitude, $9N+81N^2$, pair sector in relative coordinates & amplitude, $9N+81N^2$, relative coordinates, pair sector scaled by $\lambda$ & expectation-level, $12N$ rail qubits\\
covers & collision and streaming & collision and streaming & collision, streaming, transforms & dissipative relaxation substage\\
qubits & $\lceil\log_2(9N+81N^2)\rceil + 7$ & $4n + 10$ plus LCU ancillas & $4n + 14$ & $18N$ with reset ancillas\\
$\alpha$ of the collision block & 256 (model of \citealp{SanavioSimonSucci2025}), 171 & $\sigma_{\max}$ of the population stage: 9.4 ($\omega = 1.5$), 1.7 ($\omega = 0.2$) & $1 + \omega_{\max}/(\sqrt2\lambda)$ & none\\
$p$ per step, collision & $6\times10^{-5}$ & order $10^{-2}$ (reported, $N = 16$) & 0.82 ($\lambda=10$), 0.98 ($\lambda=100$) & 1\\
success of $T$ steps with readout & $p^T$ & $p^T$ & $e^{-2}/(a^2T^2N)$ & 1 for the dissipative block\\\bottomrule
\end{tabular}\end{table}

\section{Range of validity}\label{sec:limits}
Four mechanisms delimit the range of the present results. Level-2 truncation drops the feedback of the quadratic term on the pair sector, at $O(\kappa\Ma^3)$ per step, so it holds for times below the advective time or for flows that decay viscously before it. In the $\omega = 1.9$, $\Ma = 0.17$ test on $32\times32$ the population lift does not remain stable over 300 steps, and the moment lift, being the same operator in another basis, shows the same behavior. The norm loss of the collision stage is the relaxed non-equilibrium norm of Eq.~(\ref{eq:loss}), so it grows with the gradients that the linear evolution generates in one step, 1.3 percent on $4\times4$ at $\kappa = \pi/2$ and $4\times10^{-4}$ on $32\times32$ at the same Mach number. The near-unit success probabilities reported above accordingly apply to resolved-flow regimes. The readout amplitude of the nonlinear register, $1/(1+\lambda^2N)$, decreases as the pair-sector scaling raises the per-step probability towards one. This sets the factor $T^2N$ in Eq.~(\ref{eq:total}), and the trade-off is explicit in $\lambda$. The transforms between populations and moments require about 900 gates per step at the generic bound for a four-qubit unitary, a number independent of $N$ that a structured decomposition of the $9\times9$ orthogonal $U$ could reduce. The CPTP route of \citet{KhanSucciYao2026} keeps the dissipative substage deterministic on $O(N)$ qubits. In the hybrid pipeline described there, the nonlinear equilibrium remains to be supplied by a Carleman-type lift, and Eq.~(\ref{eq:csite}) provides one sparse representation of this component in the present basis.

\section{Conclusion}\label{sec:conclusion}
In the moment representation the level-2 Carleman hierarchy of the lattice Boltzmann equation closes under collision but not under streaming. Streaming generates cross-site terms associated with the work of pressure, advection and stress on the momentum flux, so that the full pair array remains necessary. The moment lift therefore has the same dynamics and truncation error as the population lift, but changes the structure of the operator to be block encoded. The low success probability of population-space constructions is due to the Euclidean metric of the encoding and to the density of the population basis rather than to the dissipation, which removes only the non-equilibrium norm relaxed in the step. For the D2Q9 scheme considered here, the weighted Hermite representation reduces the collision stage to a diagonal operator plus a sparse coupling, and a structured block encoding of this form raises the single-step success probability from the $10^{-5}$ to $10^{-2}$ range of population-space constructions to values close to unity for resolved flows. With the relative-coordinate storage of the pair sector \citep{BastidaZamora2026}, one time step requires $O(\log^2N)$ Toffoli gates, as verified by statevector simulation, while pair-sector scaling changes the multi-step encoding and readout overhead from exponential to polynomial in $T$ and $N$. The construction enables coherent multi-step Carleman evolution without the exponential block-encoding penalty. Future work should address three-dimensional lattices, boundary conditions, forcing and the hybrid combination with deterministic CPTP relaxation \citep{KhanSucciYao2026}.

\section*{Acknowledgements}
The research is financed by the Swedish Transport Administration in the project ``GEneric Multidisciplinary optimization for sail INstallation on wInd-assisted ships'' (GEMINI, Grant No. TRV~2023/32107).

\section*{Data availability}
The data that support the findings of this study, and the scripts that produce every number, table and figure, are available from the authors upon reasonable request.

\bibliographystyle{plainnat}
\bibliography{refs}

\begin{thebibliography}{22}
\providecommand{\natexlab}[1]{#1}
\providecommand{\url}[1]{\texttt{#1}}
\expandafter\ifx\csname urlstyle\endcsname\relax
  \providecommand{\doi}[1]{doi: #1}\else
  \providecommand{\doi}{doi: \begingroup \urlstyle{rm}\Url}\fi

\bibitem[Bastida~Zamora et~al.(2026)Bastida~Zamora, Budinski, Lahtinen, and
  Sagaut]{BastidaZamora2026}
A.~D. Bastida~Zamora, L.~Budinski, V.~Lahtinen, and P.~Sagaut.
\newblock Quantum lattice {B}oltzmann method for several time steps: a local
  {C}arleman linearization algorithm.
\newblock \emph{Physical Review E}, 113:\penalty0 035307, 2026.
\newblock \doi{10.1103/kbx3-c35d}.

\bibitem[Bharadwaj and Sreenivasan(2023)]{Bharadwaj2023}
S.~S. Bharadwaj and K.~R. Sreenivasan.
\newblock Hybrid quantum algorithms for flow problems.
\newblock \emph{Proceedings of the National Academy of Sciences}, 120:\penalty0
  e2311014120, 2023.

\bibitem[Budinski(2021)]{Budinski2021}
L.~Budinski.
\newblock Quantum algorithm for the advection-diffusion equation simulated with
  the lattice {B}oltzmann method.
\newblock \emph{Quantum Information Processing}, 20:\penalty0 57, 2021.

\bibitem[Fowler et~al.(2012)Fowler, Mariantoni, Martinis, and
  Cleland]{Fowler2012}
A.~G. Fowler, M.~Mariantoni, J.~M. Martinis, and A.~N. Cleland.
\newblock Surface codes: towards practical large-scale quantum computation.
\newblock \emph{Physical Review A}, 86:\penalty0 032324, 2012.

\bibitem[Gaitan(2020)]{Gaitan2020}
F.~Gaitan.
\newblock Finding flows of a {N}avier-{S}tokes fluid through quantum computing.
\newblock \emph{npj Quantum Information}, 6:\penalty0 61, 2020.

\bibitem[Gily{\'e}n et~al.(2019)Gily{\'e}n, Su, Low, and Wiebe]{GSLW2019}
A.~Gily{\'e}n, Y.~Su, G.~H. Low, and N.~Wiebe.
\newblock Quantum singular value transformation and beyond: exponential
  improvements for quantum matrix arithmetics.
\newblock In \emph{Proc. 51st ACM STOC}, pages 193--204, 2019.

\bibitem[Itani and Succi(2022)]{ItaniSucci2022}
W.~Itani and S.~Succi.
\newblock Analysis of {C}arleman linearization of lattice {B}oltzmann.
\newblock \emph{Fluids}, 7:\penalty0 24, 2022.

\bibitem[Jennings et~al.(2026)Jennings, Korzekwa, Lostaglio, Ashworth, Marsili,
  and Rolston]{Jennings2026}
D.~Jennings, K.~Korzekwa, M.~Lostaglio, R.~Ashworth, E.~Marsili, and
  S.~Rolston.
\newblock End-to-end quantum algorithm for nonlinear fluid dynamics with
  bounded quantum advantage.
\newblock \emph{PRX Quantum}, 2026.
\newblock \doi{10.1103/xysy-q3fp}.
\newblock in press.

\bibitem[Khan et~al.(2026)Khan, Succi, and Yao]{KhanSucciYao2026}
M.~I. Khan, S.~Succi, and H.-D. Yao.
\newblock Deterministic realization of classical dissipation on quantum
  computers.
\newblock \emph{arXiv:2604.25429}, 2026.

\bibitem[Lallemand and Luo(2000)]{LallemandLuo2000}
P.~Lallemand and L.-S. Luo.
\newblock Theory of the lattice {B}oltzmann method: dispersion, dissipation,
  isotropy, {G}alilean invariance, and stability.
\newblock \emph{Physical Review E}, 61:\penalty0 6546, 2000.

\bibitem[Liu et~al.(2021)Liu, Kolden, Krovi, Loureiro, Trivisa, and
  Childs]{Liu2021}
J.-P. Liu, H.~{\O}. Kolden, H.~K. Krovi, N.~F. Loureiro, K.~Trivisa, and A.~M.
  Childs.
\newblock Efficient quantum algorithm for dissipative nonlinear differential
  equations.
\newblock \emph{Proceedings of the National Academy of Sciences}, 118:\penalty0
  e2026805118, 2021.

\bibitem[Mezzacapo et~al.(2015)Mezzacapo, Sanz, Lamata, Egusquiza, Succi, and
  Solano]{Mezzacapo2015}
A.~Mezzacapo, M.~Sanz, L.~Lamata, I.~L. Egusquiza, S.~Succi, and E.~Solano.
\newblock Quantum simulator for transport phenomena in fluid flows.
\newblock \emph{Scientific Reports}, 5:\penalty0 13153, 2015.

\bibitem[Nielsen and Chuang(2010)]{NielsenChuang2010}
M.~A. Nielsen and I.~L. Chuang.
\newblock \emph{Quantum Computation and Quantum Information}.
\newblock Cambridge University Press, Cambridge, 10th anniversary edition,
  2010.

\bibitem[Preskill(2018)]{Preskill2018}
J.~Preskill.
\newblock Quantum computing in the {NISQ} era and beyond.
\newblock \emph{Quantum}, 2:\penalty0 79, 2018.

\bibitem[Sanavio and Succi(2024)]{SanavioSucci2024}
C.~Sanavio and S.~Succi.
\newblock Lattice {B}oltzmann-{C}arleman quantum algorithm and circuit for
  fluid flows at moderate {R}eynolds number.
\newblock \emph{AVS Quantum Science}, 6:\penalty0 023802, 2024.

\bibitem[Sanavio et~al.(2024)Sanavio, Scatamacchia, de~Falco, and
  Succi]{SanavioThreeRoutes2024}
C.~Sanavio, R.~Scatamacchia, C.~de~Falco, and S.~Succi.
\newblock Three {C}arleman routes to the quantum simulation of classical
  fluids.
\newblock \emph{Physics of Fluids}, 36:\penalty0 057143, 2024.

\bibitem[Sanavio et~al.(2025)Sanavio, Simon, Ralli, Love, and
  Succi]{SanavioSimonSucci2025}
C.~Sanavio, W.~A. Simon, A.~Ralli, P.~Love, and S.~Succi.
\newblock {C}arleman-lattice-{B}oltzmann quantum circuit with matrix access
  oracles.
\newblock \emph{Physics of Fluids}, 37:\penalty0 037123, 2025.

\bibitem[Succi(2026)]{SucciNote2026}
S.~Succi.
\newblock Proposal of a new quantum algorithm for fluid flows: multi-time
  relaxation {C}arleman lattice {B}oltzmann.
\newblock Internal note dated 18 June, 2026.

\bibitem[Succi et~al.(2023)Succi, Itani, Sreenivasan, and Steijl]{SucciEPL2023}
S.~Succi, W.~Itani, K.~Sreenivasan, and R.~Steijl.
\newblock Quantum computing for fluids: where do we stand?
\newblock \emph{EPL}, 144:\penalty0 10001, 2023.

\bibitem[Succi et~al.(2026)Succi, Sanavio, Cappelli, and
  Love]{SucciSanavioLove2026}
S.~Succi, C.~Sanavio, L.~Cappelli, and P.~Love.
\newblock The foundational value of quantum computing for classical fluids.
\newblock \emph{EPL}, 153:\penalty0 28001, 2026.

\bibitem[Todorova and Steijl(2020)]{Todorova2020}
B.~N. Todorova and R.~Steijl.
\newblock Quantum algorithm for the collisionless {B}oltzmann equation.
\newblock \emph{Journal of Computational Physics}, 409:\penalty0 109347, 2020.

\bibitem[Wang et~al.(2025)Wang, Meng, Zhao, and Yang]{Wang2025}
B.~Wang, Z.~Meng, Y.~Zhao, and Y.~Yang.
\newblock Quantum lattice {B}oltzmann method for simulating nonlinear fluid
  dynamics.
\newblock \emph{npj Quantum Information}, 11:\penalty0 196, 2025.

\end{thebibliography}
\end{document}